\documentclass[reprint,amsmath,amssymb,amsfonts,aps,showpacs]{revtex4-1}

\usepackage[normalem]{ulem}
\usepackage{graphicx}  % needed for figures
\usepackage{dcolumn}   % needed for some tables
\usepackage{bm}        % for math
\usepackage{amssymb}   % for math
\usepackage{float}
\usepackage{comment}
\usepackage{braket}
\usepackage{scalerel}
\usepackage{xcolor}
\begin{document}

\title{Correlating photons using the collective nonlinear response of atoms weakly coupled to an optical mode}

\author{Adarsh S. Prasad$^1$, Jakob Hinney$^1$,	Sahand Mahmoodian$^{2,\ast}$, Klemens Hammerer$^2$, Samuel Rind$^1$, Philipp Schneeweiss$^{1,3}$, Anders S. S\o rensen$^4$, J\"urgen Volz$^{1,3}$, and Arno Rauschenbeutel$^{1,3,\dagger}$}
\affiliation{$^1$Vienna Center for Quantum Science and Technology, TU Wien-Atominstitut, Stadionallee 2, 1020 Vienna, Austria}
\affiliation{$^2$Institute for Theoretical Physics, Institute for Gravitational Physics (Albert Einstein Institute), Leibniz University Hannover, Appelstra{\ss}e 2, 30167 Hannover, Germany}
\affiliation{$^3$Department of Physics, Humboldt-Universität zu Berlin, 10099 Berlin, Germany}
\affiliation{$^4$Center for Hybrid Quantum Networks (Hy-Q), Niels Bohr Institute, University of Copenhagen, Blegdamsvej 17, DK-2100 Copenhagen, Denmark}

\date{\today}

\begin{abstract}
Photons in a nonlinear medium can repel or attract each other, resulting in a strongly correlated quantum many-body system~\cite{Chang2014NPHOT,Chang2018RMP}. Typically, such strongly correlated states of light arise from the extreme nonlinearity granted by quantum emitters that are strongly coupled to a photonic mode~\cite{Lodahl2015RMP, Chang2018RMP}. However, in these approaches, unavoidable dissipation, like photon loss, blurs nonlinear quantum effects. Here, we generate strongly correlated photon states using only weak coupling and taking advantage of dissipation. We launch light through an ensemble of non-interacting waveguide-coupled atoms, which induce correlations between simultaneously arriving photons through collectively enhanced nonlinear interactions. These correlated photons then experience less dissipation than the uncorrelated ones. Depending on the number of atoms, we experimentally observe strong photon bunching or anti-bunching of the transmitted light. This realization of a collectively enhanced nonlinearity may turn out transformational for quantum information science and opens new avenues for generating nonclassical light, covering frequencies from the microwave to the X-ray regime.
\end{abstract}

\maketitle

Photons that strongly interact via a quantum nonlinear medium  exhibit complex out-of-equilibrium quantum many-body dynamics which may enable one to tailor and control the photon statistics of the light~\cite{Chang2014NPHOT,Chang2018RMP}. The resulting quantum correlated light can then act as a key resource in quantum sensing, quantum metrology, quantum communication, as well as quantum simulation and information processing. Recently, significant advances have been made in mediating interactions between optical photons by strongly coupling them to quantum emitters and exploiting the inherently nonlinear response of the latter~\cite{Lodahl2015RMP, Chang2018RMP}. A number of methods have been used for this purpose, such as resonant enhancement via high finesse optical cavities \cite{Birnbaum2005Nature,Dayan08,Faraon2008NPHYS, Reinhard2012NPHOT, Reiserer2014Nature, Volz2014, Hamsen2017PRL}, 
collective response of strongly interacting Rydberg atoms \cite{Dudin2012Science, Parigi2012PRL,  Maxwell2013PRL,Firstenberg2013Nature, Baur2014PRL, Thompson2017Nature, Stiesdal2018PRL, Tiarks2019NPHYS} or efficient coupling of single quantum emitters to waveguides \cite{Goban2014NCOM, Javadi2015NCOM, Coles2016NCOM,Tuerschmann2019}. 
However, the implementation of strong interactions between individual optical photons remains a challenging goal.
In particular, such approaches are often significantly impaired by unavoidable dissipative processes which cause photon loss and blur nonlinear quantum effects. 

Here, we experimentally demonstrate a novel mechanism where a strongly dissipative nonlinear medium, consisting of weakly coupled quantum emitters, is harnessed to generate strongly correlated states of light~\cite{Mahmoodian2018PRL}. Specifically, we launch a weak resonant laser light field through an ensemble of non-interacting weakly coupled atoms and analyze the second order correlation function $g^{(2)}(\tau)$ of the transmitted light. Adjusting the number of atoms, we continuously change the photon statistics from antibunching down to $g^{(2)}(0)=0.41\pm 0.09$ to bunching of up to $g^{(2)}(0)=22\pm 5$. This demonstrates, for the first time, coherent collective enhancement of photon---photon interactions in an ensemble of otherwise non-interacting emitters. Consequently, our scheme may turn out transformational in quantum information science. For example, it offers a fundamentally new approach to realizing single photon sources which may outperform sources based on single quantum emitters with comparable coupling strength.

\begin{figure}
	\includegraphics[width=\columnwidth]{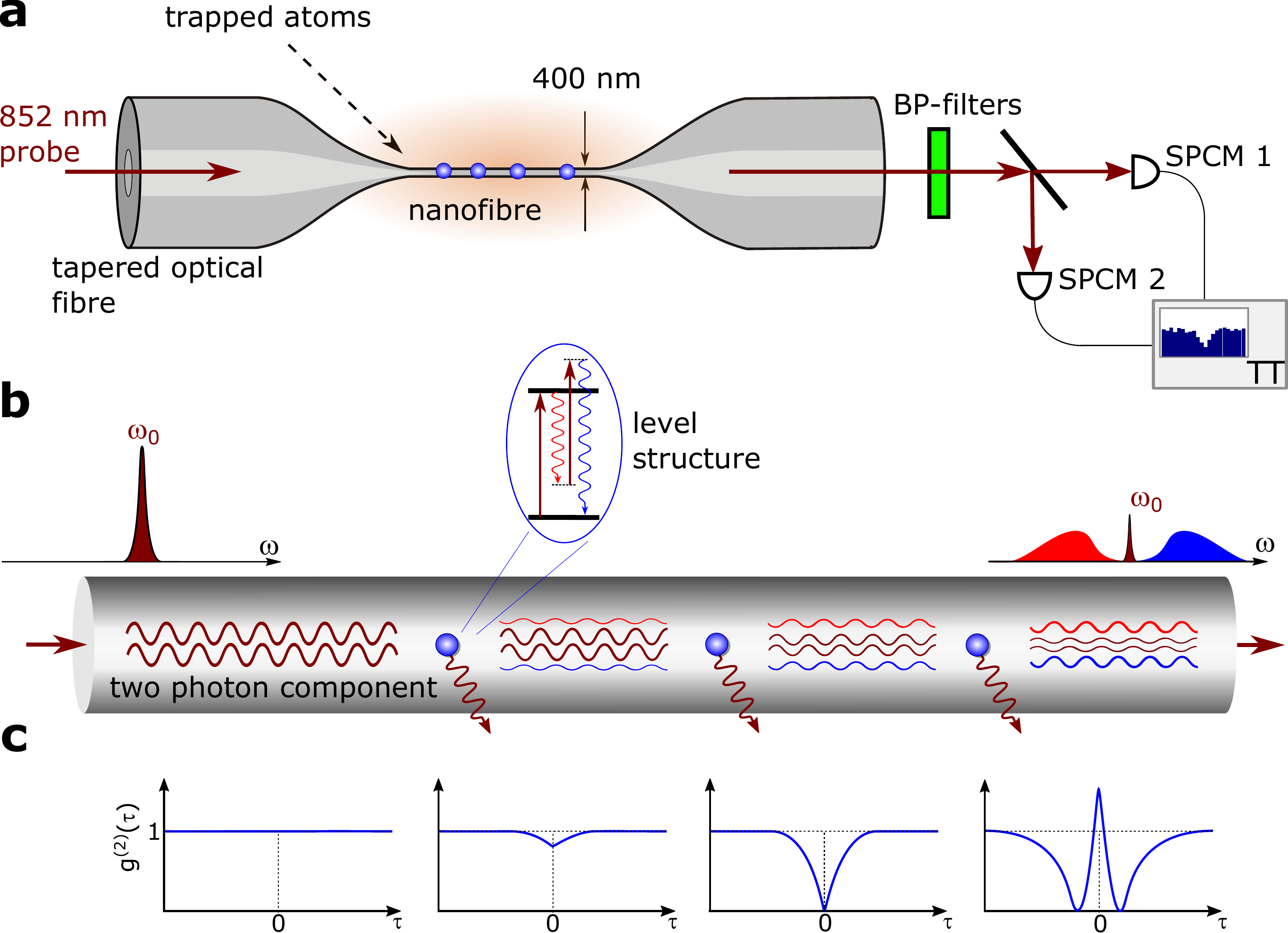}
	\caption{ \textbf{Schematics of the experimental process. a}  Laser-cooled Cs atoms are trapped along an optical nanofibre which is realized as the waist of a tapered optical fibre. Laser light (wavelength: 852 nm), resonant with the Cs D2 cycling transition, is launched into the fibre and interacts with the atoms. After passing a set of band-pass (BP) filters, the photon statistics of the transmitted light is analyzed using a Hanbury-Brown-Twiss (HBT) setup consisting of a beam splitter with a single-photon counting module (SPCM) in each output port and a timetagging unit (TT) for recording the photon arrival times. 
	\textbf{b} Probe photons that interact one by one with an atom in the evanescent field of the nanofibre are scattered out of the optical mode and are subject to exponential losses. However, when two photons simultaneously interact with an atom, the scattering process generates energy-time entanglement between the photons (see inset). As a consequence, the spectral distribution of the scattered two-photon component is broader and, therefore, subject to reduced absorption by the rest of the ensemble. The scattering amplitudes from different atoms interfere constructively resulting in collective enhancement of the two-photon scattering process. The scattered two-photon component is phase shifted by $\pi$ with respect to the unscattered one. For ensembles of intermediate length, destructive interference between the scattered and the unscattered two-photon component therefore results in anti-bunched photon statistics. For a sufficiently long ensemble, all single photons are lost and only the scattered two photon components survives. In this situation, the photon statistics shows strong bunching and the spectrum of the transmitted light is dominated by a red and a blue sideband. 
	\textbf{c} Second-order correlation function of the light expected for different positions along the atomic ensemble. 
}\label{fig:setup}
\end{figure}

The experimental setup is schematically depicted in Fig.~\ref{fig:setup}\textbf{a}. An ensemble of laser-cooled cesium (Cs) atoms is trapped in a potential consisting of two diametric linear arrays of individual trapping sites along a 400-nm diameter nanofibre, located at a distance of $\sim$ 250 nm from the fibre surface. Each site contains at most one atom and offers subwavelength confinement of the atoms in all three spatial dimensions \cite{Vet10}; see Methods for further details.
We send probe light through the nanofibre that couples to the atoms via the evanescent field and is resonant with the cycling transition of the Cs D2 line  ($6S_{1/2}, F = 4  \rightarrow  6P_{3/2}, F^\prime = 5$, $\lambda=852$ nm). The input power of the light is $P_{\rm in}=2.35$~pW which corresponds to a saturation parameter of $S_0=P_{\rm in}/P_{\rm sat}=0.02$ 
where $P_{\rm sat}$ is the power required to obtain saturation intensity at the trapping sites. The light is quasi-linearly polarised, and the polarisation axis is aligned such that we realize chiral light-matter coupling \cite{Petersen2014Science,Mitsch2014NCOM,Lodahl2017Nature}, where the atoms dominantly interact with light in the forward propagating mode and the backscattering is strongly suppressed. The coupling strength of the atoms to the forward propagating nanofibre mode $\beta=\Gamma_{\rm fw}/\Gamma$ is defined as the ratio of the spontaneous emission rate of an atom in the forward direction of the waveguide, $\Gamma_{\rm fw}$, and the total spontaneous emission rate into all channels, $\Gamma$.

\begin{figure*}[htb]
	\includegraphics[width=0.9\textwidth]{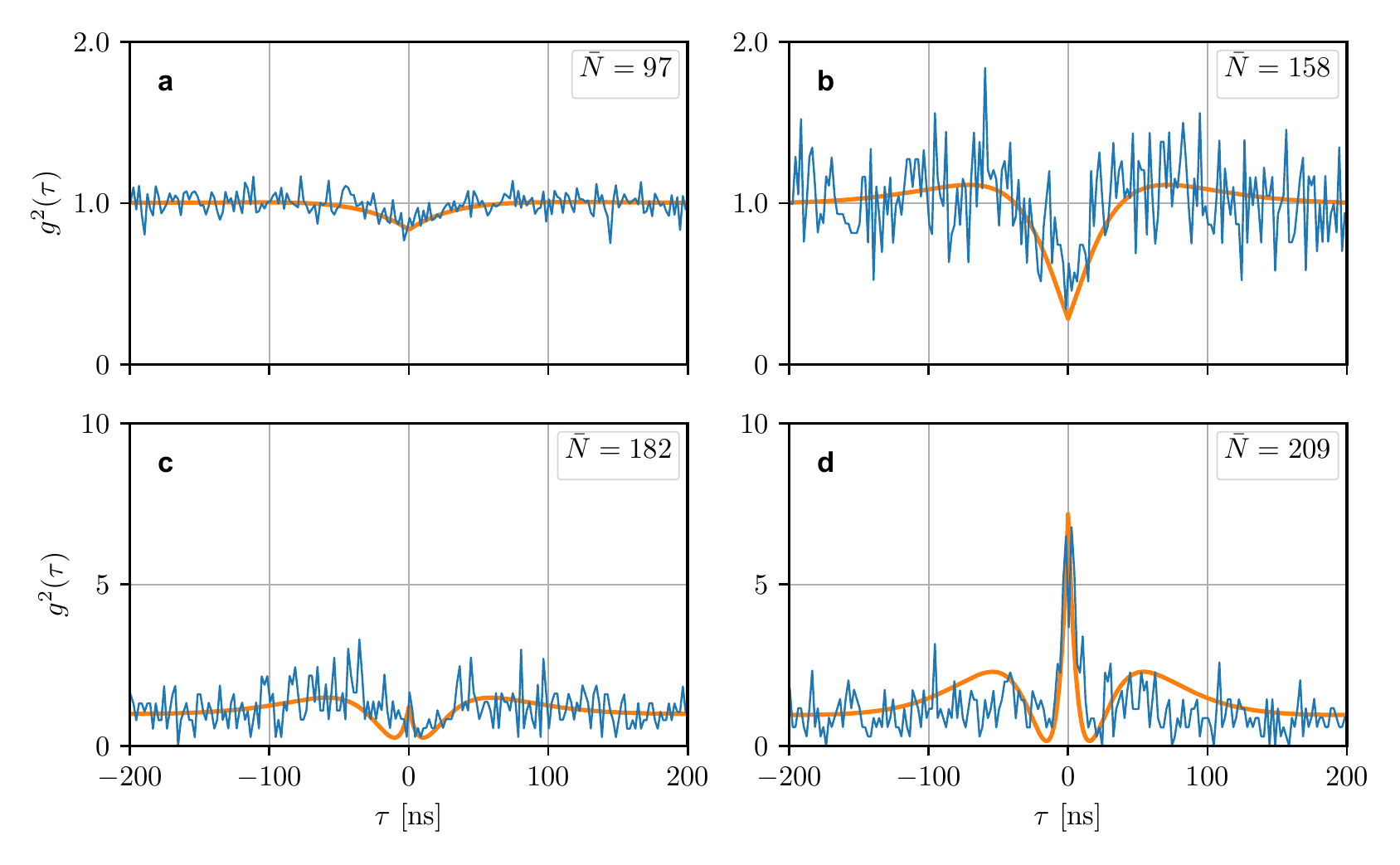}
	\caption{\label{fig_g2t_examples} \textbf{Measured second order correlation functions for four different mean atom numbers.} The blue line is the experimental data (2 ns binning) and the orange line is the theory prediction for our experimental parameters (see main text). The measured optical depth for the four panels are \textbf{a:} OD$=3.15$, \textbf{b:} OD$=5.13$,  \textbf{c:} OD$=5.88$, and  \textbf{d:} OD$=6.75$. The same data sets were used to compute $g^{(2)}(0)$ for the corresponding ODs in Fig.~\ref{fig_g2vsN}. \textbf{a} For small atom numbers $\bar N=97$, we observe antibunching which increases with increasing atom number. \textbf{b} The antibunching reaches its maximum at $\bar{N}=158$ atoms. \textbf{c} When further increasing the atom number to $\bar N=182$, $g^{(2)}(\tau)$ starts to exhibit a peak at $\tau = 0$ that turns to strong photon bunching for very large atom numbers ($\bar N=209$) \textbf{d}. The oscillatory behavior of $g^{(2)}(\tau)$ originates from the quantum beat of the scattered and unscattered two-photon component.}
\end{figure*}

In the linear optical regime, the atoms act as a narrow-band spectral filter and induce strong (exponential) attenuation of individually propagating resonant probe photons. When two probe photons are incident simultaneously, however, the nonlinear interaction with each atom also induces energy-time entanglement between the photons. As a result, each of the two scattered photons then features a broadened spectrum with frequency components that are red- and blue-detuned from the atomic resonance. The scattered photons thus experience reduced (subexponential) propagation loss~\cite{Mahmoodian2018PRL}, see Fig.~\ref{fig:setup}\textbf{b}. Importantly, the amplitudes of the correlated photon pairs that arise from scattering by individual atoms add up coherently, thus giving rise to a collective enhancement of the process. The strength of these linear and nonlinear processes together with the number of atoms coupled to the optical mode defines the ratio of the scattered and unscattered one- and two-photon components in the output mode. 

We investigate the photon statistics of the transmitted probe light by sending it onto a Hanbury-Brown-Twiss setup consisting of a 50/50 beamsplitter and a single photon counting module (SPCM) in each output. The normalized histogram of the time differences, $\tau$, between the photon detection events then yields the second order correlation function, $g^{(2)}(\tau)$, of the transmitted light, see Methods. In addition, we use the measured count rates from the SPCMs for determining the transmitted power through the nanofibre. This allows us to infer the optical depth (OD) of the trapped ensemble for each of the $2.6\times 10^6$ experimental runs. After sorting the data according to the measured OD and averaging the data in each OD interval, we obtain a set of 54 second order correlation functions; see Fig.~S2 in the Supplementary Information. 

\begin{figure}[h]
	\includegraphics[width=\columnwidth]{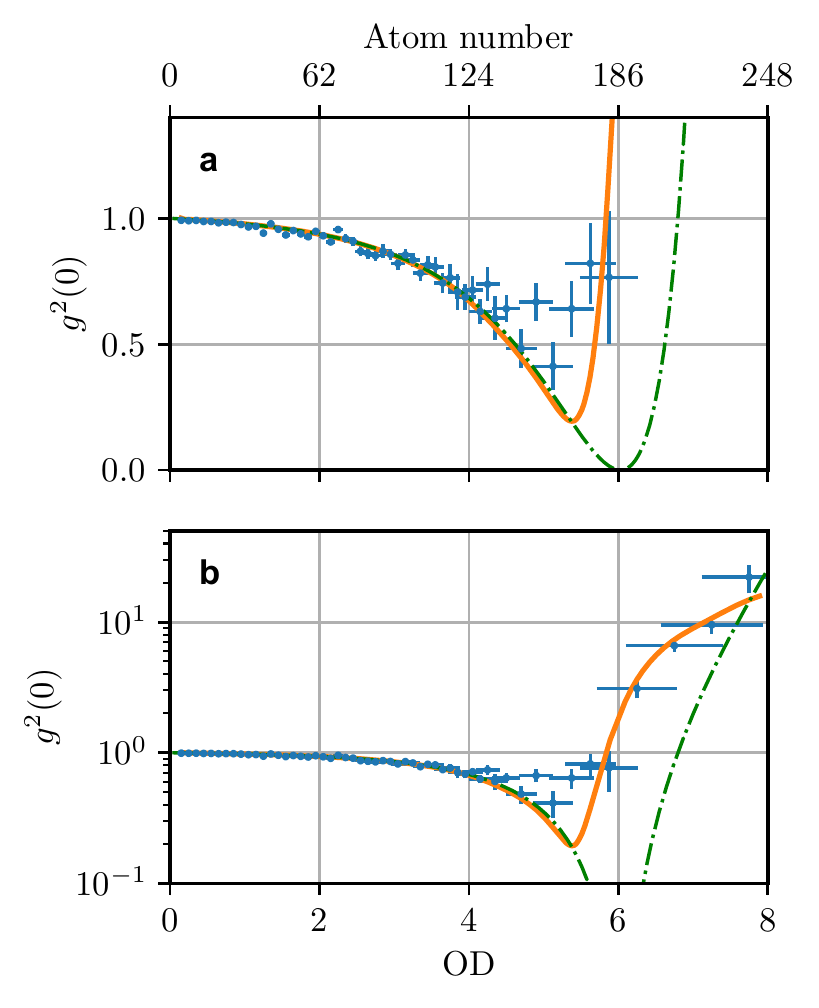}
	\caption{\label{fig_g2vsN} \textbf{Correlations at zero time delay  vs. number of trapped atoms} The blue data points show the zero time delay value $g^2(0)$ of the measured second order correlation functions as a function of the optical depth of the atomic ensemble (lower x-axis) or average number of trapped atoms (upper x-axis). For better visibility, we plot the same data using a linear (\textbf{a}) and a logarithmic (\textbf{b}) scale for the y-axis. The values $g^{(2)}( 0)$ and their errors are obtained from maximum likelihood fits to the individual correlation functions, see Methods. The solid red line is the theory prediction taking into account the experimental uncertainty in OD estimation with the coupling strength $\beta$ as only fit parameter, see Methods. For comparison, we also show the theory curve without uncertainty in atom number for the same value of $\beta$ (dashed green curve). The number of trapped atoms on the $x$-axis is determined from the measured OD using the measured value of $\beta$. The error bars in $x$-direction indicate the spread in atom  numbers that enter in the measured correlation function, see Methods. }
\end{figure}

Figures~\ref{fig_g2t_examples} and \ref{fig_g2vsN} summarize the main results of these correlation measurements: Figure~\ref{fig_g2t_examples} exemplarily shows four second order correlation functions for different regimes and Fig.~\ref{fig_g2vsN} shows the value of $g^{(2)}(\tau=0)$ as a function of the OD for all measured correlation functions. The solid orange curves in both figures are theory predictions \cite{Mahmoodian2018PRL} taking into account the experimental variation in the atom number distribution. The assumed coupling strength is derived from fitting the model to the data in Fig.~\ref{fig_g2vsN} with $\beta$ as the only fit parameter; see Methods. The fitted value of $\beta=0.81\% \pm0.02\%$ agrees well with the value of $\beta=0.83\%\pm0.03\%$, derived from an independent saturation measurement; see Methods. The resulting theory curves also agree well with the experimental data in Figs.~\ref{fig_g2t_examples} and \ref{fig_g2vsN} and Fig.~S2 in the Supplementary Information.

For vanishing OD, we observe a flat $g^{(2)}(\tau)$ with $g^{(2)}(0)\approx1$, as expected for the photon statistics of our probe laser. With increasing OD, one observes photon anti-bunching, i.e., $g^{(2)}(0)$ starts to fall below one; see Fig.~\ref{fig_g2t_examples}\textbf{a}. It originates from destructive quantum interference of the unscattered two-photon component and the correlated two-photon component. This scattered component arises from the nonlinear interaction of the probe light with the atoms and is shifted by $\pi$ with respect to the phase of the unscattered two-photon component~\cite{Mahmoodian2018PRL}, thereby enabling the destructive interference. As a result, the overall two-photon amplitude at zero delay is reduced which then leads to a reduction of $g^{(2)}(0)$. For larger time differences, $\tau$, the frequency difference of the two photons of the scattered two photon component from the laser frequency gives rise to an oscillatory behavior of $g^{(2)}(\tau)$~\cite{Legero2004PRL}. With increasing OD, the ratio of scattered to unscattered two-photon component grows and the photon anti-bunching gets more pronounced. For an OD of 5.13, corresponding to a mean number of atoms of $\bar{N}=158$, the anti-bunching reaches its smallest value of $g^{(2)}(0)=0.41 \pm 0.09$, see Fig.~\ref{fig_g2t_examples}\textbf{b}. Ideally, the theory even predicts perfect anti-bunching when the ratio of  scattered to unscattered two-photon amplitude is similar; see dashed line in Fig.~\ref{fig_g2vsN}. 
The reduced contrast of the measured correlation functions stems to the largest part from the fact that each correlation function is averaged over a spread of ODs. This is on the one hand due to the finite binning of ODs and on the other hand because photon shot noise impairs the precise determination of the OD from the transmission in the individual experimental runs.

When further increasing the OD or the number of trapped atoms, the ratio of scattered to unscattered photon pairs increases further. As a consequence, $g^{(2)}(\tau)$ starts to exhibit a peak at $\tau = 0$ and perfect anticorrelations are expected for finite time delays $\tau\neq0$; see Fig.~\ref{fig_g2t_examples}\textbf{c}. For very large ODs, the scattered two-photon-component dominates in the output field.  This manifests in strong photon-bunching, as observed in our measurements for mean atom numbers exceeding $\bar{N}\approx180$; see Figs.~\ref{fig_g2t_examples}\textbf{d} and \ref{fig_g2vsN}. In the extremely high OD-limit, all transmitted photons originate from nonlinear interaction with the atoms and the bunching increases indefinitely. 

Our results show that ensembles of weakly coupled atoms can be used to realize strongly correlated many-body states of photons. In contrast to previous approaches that rely on strong light--matter coupling and suppression of dissipation, the underlying dynamics of our non-equilibrium many-body quantum system is based on an interplay of weak optical nonlinearities, collective enhancement, and finite dissipation. It is well-known that collective enhancement can be used to compensate for a limited coupling efficiency and enable strong coupling with ensembles of otherwise non-interacting atoms~\cite{dua01,Ham10}. The generation of non-trivial field states have so far, however, relied on strong optical driving fields to enhance the optical non-linearity. In contrast, no-such driving fields are used in this experiment. The present approach, thus, extends the concept of collective enhancement also to direct photon-photon interactions.
This  broadens the range of  possible applications, in particular in the realm of quantum information science.    

As an example, consider employing the observed antibunching for the generation of single photons. For small coupling strengths ($\beta<0.1$) and low input photon rates ($n_{in}\lesssim 0.1 \Gamma/\beta$) we can approximate the power transmission at which we obtain perfect anti-bunching by $T\approx\beta$; see Methods. 
Thus, it is possible to realize a stream of anti-bunched light with a photon rate $n_{\rm out}\approx0.1\Gamma$ in the output mode, independent of the type of emitter and their coupling strength $\beta$ to the optical mode. Surprisingly, this is much larger than the maximum photon rate $\beta\cdot\Gamma/2$ that is achievable with a single photon source based on a single quantum emitter with the same coupling strength $\beta$. 
In addition, this principle is independent of the type of emitter or specific optical mode or frequency used and can be achieved for all frequencies spanning the electromagnetic spectrum without the need for precisely controlling individual emitters and their coupling strengths. 
These features makes the observed effects highly promising for realising new sources of non-classical light, such as single photon sources, in particular for wavelengths where it is not possible to achieve strong coupling of individual atoms or emitters.

\section*{Acknowledgements}
\noindent
The authors gratefully acknowledge financial support by the European Comission under the projects ErBeStA (No.800942) and the ERC  grant  NanoQuaNt, by the Austrian Science Fund (DK CoQuS Project No. W 1210-N16), by the DFG through CRC 1227 DQ-mat (project A06) and by the Danish National Research Foundation (Center of Excellence Hy-Q).

\section*{Author contributions}
\noindent 
K.H., S.M., and A.S. made the theory predictions and were responsible for modeling the data. J.H., A.P., A.R. S.R., P.S., and J.V. contributed to the design and the setting-up of the experiment. A.P. and J.H. performed the experiment. A.P. together with S.M. and J.V. was responsible for analysing the data. All authors contributed to the writing of the manuscript.

\section*{Competing financial interests}
\noindent The authors declare no competing financial interests.

\bibliography{citations_new}

\clearpage

\section*{Methods}

\noindent \textbf{Nanofibre based optical dipole trap:} We trap laser-cooled cesium (Cs) atoms using a nanofibre-based two-color optical dipole trap~\cite{Fam04,Vet10}. The repulsive blue-detuned light field has a wavelength of $\lambda=685$~nm and a power of $\sim 14$~mW, and is launched into the fibre in a running-wave configuration. A pair of counter-propagating red-detuned fibre-guided light fields is also launched into the fibre, thereby forming a standing wave. This light forms an attractive potential and has a wavelength of $\lambda=935$~nm and a power of $\sim 0.18$~mW per beam. All light fields are quasi-linearly polarised. The polarisation planes of the two red-detuned light fields are parallel to each other while the polarisation plane of the blue-detuned trapping light field is perpendicular to that. In this configuration, the minima of the optical trap potential are located at a distance of $\sim$ 250 nm from the surface of the 400-nm diameter nanofibre. The wavelengths of the trapping light fields correspond to the magic wavelengths of the Cs D2 line \cite{Gob12} which minimizes the light shifts of this optical transition of the atoms in the nanofibre-based dipole trap. 

%of around 20$\mu$K
\noindent \textbf{Ensemble preparation:} The atoms are loaded into the optical dipole trap from a cigar-shaped cloud of cold Cs atoms. The atom cloud is created using a magneto-optical trap (MOT) with elongated magnetic coils. The MOT is followed by a molasses cooling stage where the atoms are cooled down to sub-Doppler temperature and loaded into the trap. After the first molasses, the cooling light field and the MOT magnetic fields are turned off, and the red-trap power is adiabatically ramped up over a time of 10 $\mu$s. This moves the potential minimum of the trap closer to the fibre surface and results in a larger coupling strength between the trapped atom and a nanofibre-guided light field. After the power ramp of the red-detuned light field, a second molasses phase cools the remaining trapped atoms. The OD exhibited by the trapped atoms is characterized by a standard frequency sweep.  These steps are illustrated in Fig.~4, i.-iv. in the Supplementary Information. 

\noindent \textbf{Probing sequence:} The probe light field is resonant with the D2 ($6S_{1/2}, F = 4  \rightarrow  6P_{3/2}, F^\prime = 5$) transition and launched as a travelling wave into the nanofibre. The probe's polarisation is quasi-linearly polarised in the plane of the trapped atoms, i.e., it coincides with that of the red-detuned trapping light fields. The probing sequence comprises 350 resonant pulses, each with a duration of 10 $\mu$s. The pulses have a power of 2.35~pW, which corresponds to a on-resonance saturation parameter ($S_0$) of 0.02 for the first atom the laser interacts with.
The probe pulse duration is chosen such that the average kinetic energy that is transferred on the first atoms in the chain due to photon recoil is less than half the trap depth. To compensate for the heating, we apply molasses cooling with a duration of 200~$\mu$s  between two probing pulses. The total number of probe pulses and the duration of the interleaved cooling pulses are chosen such that the change of OD over the entire sequence is as small as possible.

The probe sequence is followed by an additional frequency scan to measure the OD in order to check for atoms loss (vi). Afterwards, the atoms are removed from the trap (vii) and we perform calibration measurements on the nanofibre transmission (viii), see Figure 4 in the Supplementary Information.

\noindent \textbf{Detection:} The light transmitted through the nanofibre-coupled atomic ensemble is, first, sent to a spectral filtering stage consisting of a Fabry-P\'erot cavity (spectral width: $\sim 100$ MHz) and a volume Bragg grating. In this way, the signal light can be separated from the trapping light fields and from the Raman scattering they produce when propagating in the fibre. Afterwards, the light is sent onto a Hanbury-Brown-Twiss setup that consists of a beam splitter with a single photon counting module in each output. The modules are connected to a FPGA-based time tagger which records the photon detection events with a timing resolution better than one nanosecond. After histogramming the time differences between the photon detection events in the two detectors, we can infer the second-order correlation function of the transmitted light from the data:
\begin{equation}
g^2(\tau)=\frac{\langle \hat a^\dagger(t)\hat a^\dagger(t+\tau)\hat a(t+\tau)\hat a(t)\rangle}{\langle \hat a^\dagger(t)\hat a(t)\rangle^2}
\end{equation}
Here, $\hat a(t)$ and $\hat a^\dagger(t)$ are the photon annihilation and creation operators, respectively.

\noindent \textbf{Data processing:} The FPGA records time tags for each detection event in each SPCM. We aim to minimize the effect of transient transmission signals at the beginning and end of each probe pulse, originating, e.g., from the finite rise and fall time that the acousto-optical modulator that controls the optical pulse. Thus, we only take detection event that fall into the probing interval between $1\;\mu$s and $9\;\mu$s for each probe pulse into account. Before each probe pulse train, at the beginning of each individual run, we also perform a measurement of the OD of the atomic ensemble. This increases the temperature of the atoms, and we observe an increased transmission for the first probe pulses that decays to a steady state after about $\sim15$ probing-cooling iterations. In order to have a well defined number of atoms and a precise atom-light coupling strength, we thus discard the data from the first 20 probe pulses of each experimental run. 

From the timestamps recorded for a single run, we histogram all the time differences between the photon detection events in the two detectors to obtain the coincidence histogram $c_i(\tau_i)$, where $\tau_i$ represents the time delay between the two clicks. The binsize is 2~ns, and $c_i$ represents the number of coincidences detected at this time delay. We accumulate the coincidence statistics from each run into one out of 54 histograms based on the OD$=-ln[T]$ observed in the individual run, where $T$ is the average transmission measured over the run. The resulting histograms are normalized by setting the correlation function to one for long-time delays ($\tau>200$~ns). 
The OD windows used in Figs.~\ref{fig_g2vsN} and 5 in the Supplementary Information are chosen such that increasing OD the bin-size also increases in order to compensate for the reduced signal to noise ratio in transmission. The windows corresponds to an OD-bin-size of 0.1, 0.25 and 0.5 for ODs in the lying in the intervals [0, 4.0], [4, 5] and [5, 8], respectively.

\noindent \textbf{Maximum-likelihood estimation of $g^{(2)}(0)$:} 
To obtain a good estimate the zero time delay value $g^{(2)}(0)$ from the individual second order correlation measurements, we perform a heuristic fit of the measured correlation functions in a small time window around $\tau=0$. Due to the low coincidence count rates in the data for higher ODs a standard fitting algorithms is in general not reliable. To circumvent this, we instead use a maximum likelihood estimation method (MLE) that searches for the theory $g^{(2)}_{\rm theory}(\tau)$ function that has the highest probability to reproduce the measured correlation function. As the temporal distribution of the coincidences is directly proportional to the expected second order correlation function $g^{(2)}_{\rm exp}(\tau)$, this method boils down to maximizing the probability
\begin{equation}
\mathcal{L}\left[\Theta\right] = \prod_{i=1}^{n} (g^{(2)}_{\rm theory}(\tau_i,\Theta))^{c_i}
\end{equation}
to observe the measured data assuming a theoretical correlation function $g^{(2)}_{\rm theory}(\tau,\Theta)$ by varying the parameters of the parameter set $\Theta$.
Here, $c_i(\tau_i)$ represents the values of the measured coincidence histogram. Since we are mainly interested in the zero value of the measured correlation functions, we can approximate them for small time delays by $g^{(2)}(\tau)=1-A \exp(-\Gamma |\tau|)$, where $A$ and $\Gamma$ are the two fit parameter. For optimizing the accuracy of the estimation of $g^{(2)}(0)$, we limit the fit region to the characteristic time scale of the expected correlation functions, i.e. to $|\Delta\tau|=1/\Gamma\approx 30$ ns for the region where we observe antibunching and to $|\Delta \tau|\approx15$ ns for the datasets with $\textrm{OD}>6$ where $g^{(2)}(0)>1$, see Fig.~\ref{fig_g2vsN}.

\noindent\textbf{Error estimation of $g^{(2)}(0)$:}
We estimate the error in the value of $g^{(2)}(0)$ by means of a bootstrapping method. For each datapoint in Fig.~\ref{fig_g2vsN}, we randomly generate 50 correlation functions with the same photon statistics, using the results $A$ and $\Gamma$ of the MLE fit to the measured correlation functions. For each of these samples we again perform a MLE which yield the new fit results $A'$ and $\Gamma'$. From the resulting distribution of $A'$ we use its standard deviation as estimation of the error of $A$. These standard deviations define the errorbars along the $y$-axis in the Fig.~\ref{fig_g2vsN}.

\noindent \textbf{Theory prediction of $g^{(2)}(0)$ for our experiment:} 
To model the behavior of the transmitted light as function of atom number shown in Fig.~\ref{fig_g2vsN} we first estimate the atom number distribution that enters in each data point due to our post selection on transmission. This distribution is estimated taking into account the measured probability of preparing an atomic ensemble with certain OD as well as the uncertainty in OD estimation that originates from photon shot noise in the measured transmission. The errorbar in $x-$direction in Fig.~\ref{fig_g2vsN} indicates the OD spread corresponding to the standard deviation of the atom number distribution. Taking into account this averaging in the theory \cite{Mahmoodian2018PRL} we fit the averaged second-order correlation function to the data set in Fig.~\ref{fig_g2vsN} using the coupling strength $\beta$ as the only fit parameter.

\noindent \textbf{Estimation of coupling strength ($\beta$)}: In order to estimate the coupling strength $\beta$ of a single atom to the waveguide, we carry out a saturation measurement as described in Ref.~\cite{Vet10, Rei14, Gouraud2015}. 
To do so, we send into the nanofibre a probe pulse with a duration of 10~$\mu$s with varying power, and record the optical power transmitted through the ensemble. We then fit the absorbed power versus the input power using the generalized Beer-Lambert law with $\beta$ as fit parameter. From this, we find $\beta=0.0083 \pm 0.0003$.

\noindent\textbf{Single photon output rate}: In order to get an estimate of the photon output rate that can be reached at the point of perfect antibunching, we calculated the power transmission $T$ for the case of low input photon rate of $n_{\rm in}=0.1\times \Gamma/\beta$.
For $\beta<0.1$ this transmission follows approximately $T\approx\beta^{1.17}$. Consequently, the rate of antibunched photons at the output of the ensemble is given by $n_{\rm out}=Tn_{\rm in}\approx 0.1\times\Gamma\beta^{0.17}\approx0.1\times\Gamma$. In this regime, after the interaction with the atoms, one obtains a stream of antibunched light with a rate of $0.1\Gamma$. For comparison, for a single quantum emitter-based single photon source with the same emitter-waveguide coupling strength one expects a maximum photon rate of $\beta\Gamma/2$. \\

\clearpage

\begin{figure*}[htb]
	\includegraphics[width=0.8\textwidth]{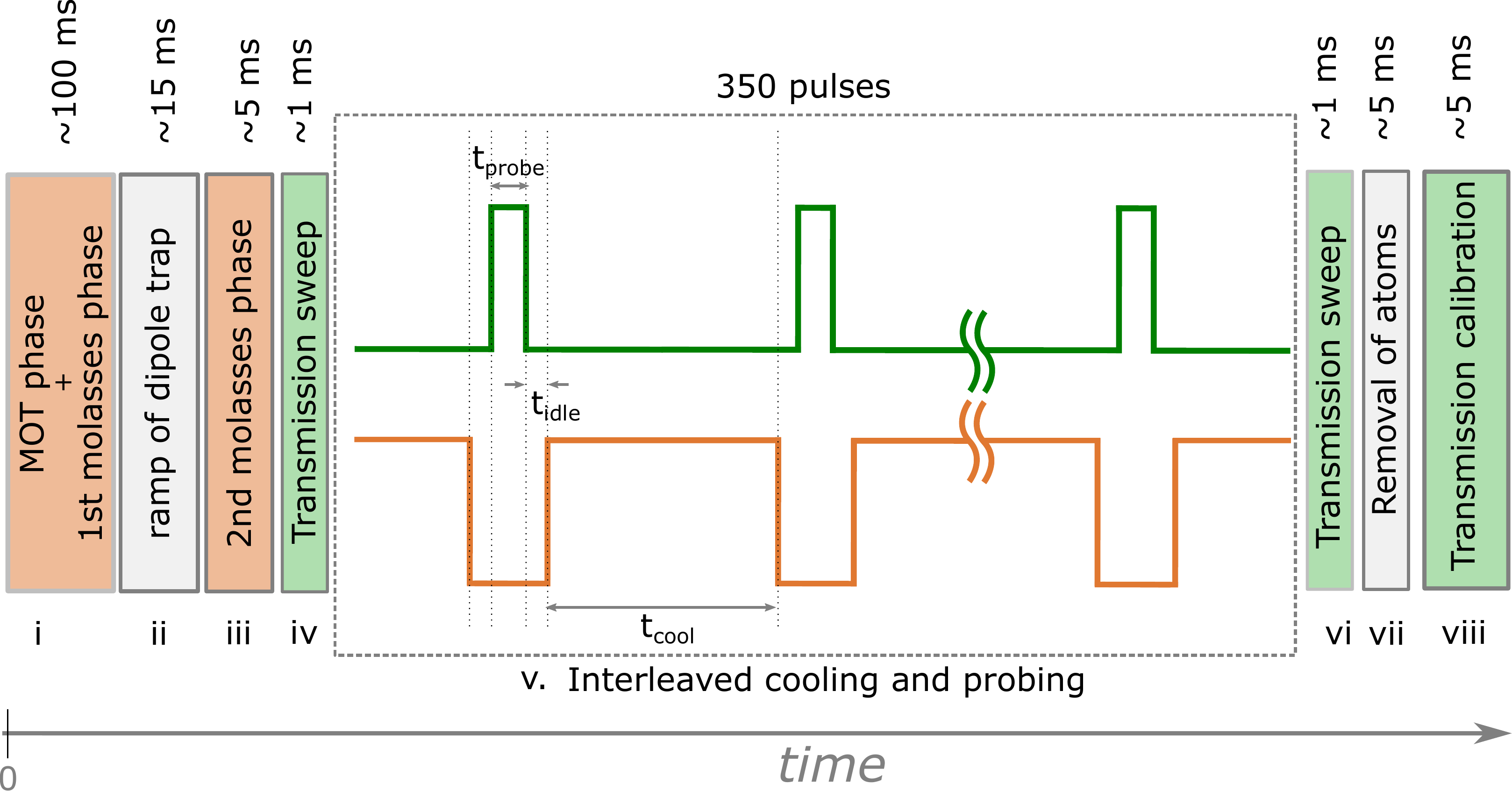}
	\caption{\textbf{Experimental sequence} The experimental sequence consists of different stages: first, a MOT and optical molasses are used to load an ensemble of cold atoms into the nanofibre-based trap (i). This is followed by a 10 ms power ramp of the red-detuned trapping field in order to increase $\beta$ while the atoms from the MOT disperse (ii). A second molasses phase is used to cool the atoms inside the trapping potential in order to compensate the heating during step-ii (iii). We then sweep the probe laser frequency across the atomic resonance and measure the transmission in order to obtain the OD of the ensemble (iv). In the main experimental sequence, we alternate 350 times between measuring transmission and re-cooling the atoms (v). After the main sequence, we again measure the OD to check if atoms were lost (vi), remove the atoms from the trap (vii), and calibrate the overall transmission through the nanofibre (viii).}
\end{figure*}

\begin{figure*}[htb]
    \includegraphics[width=0.95\textwidth]{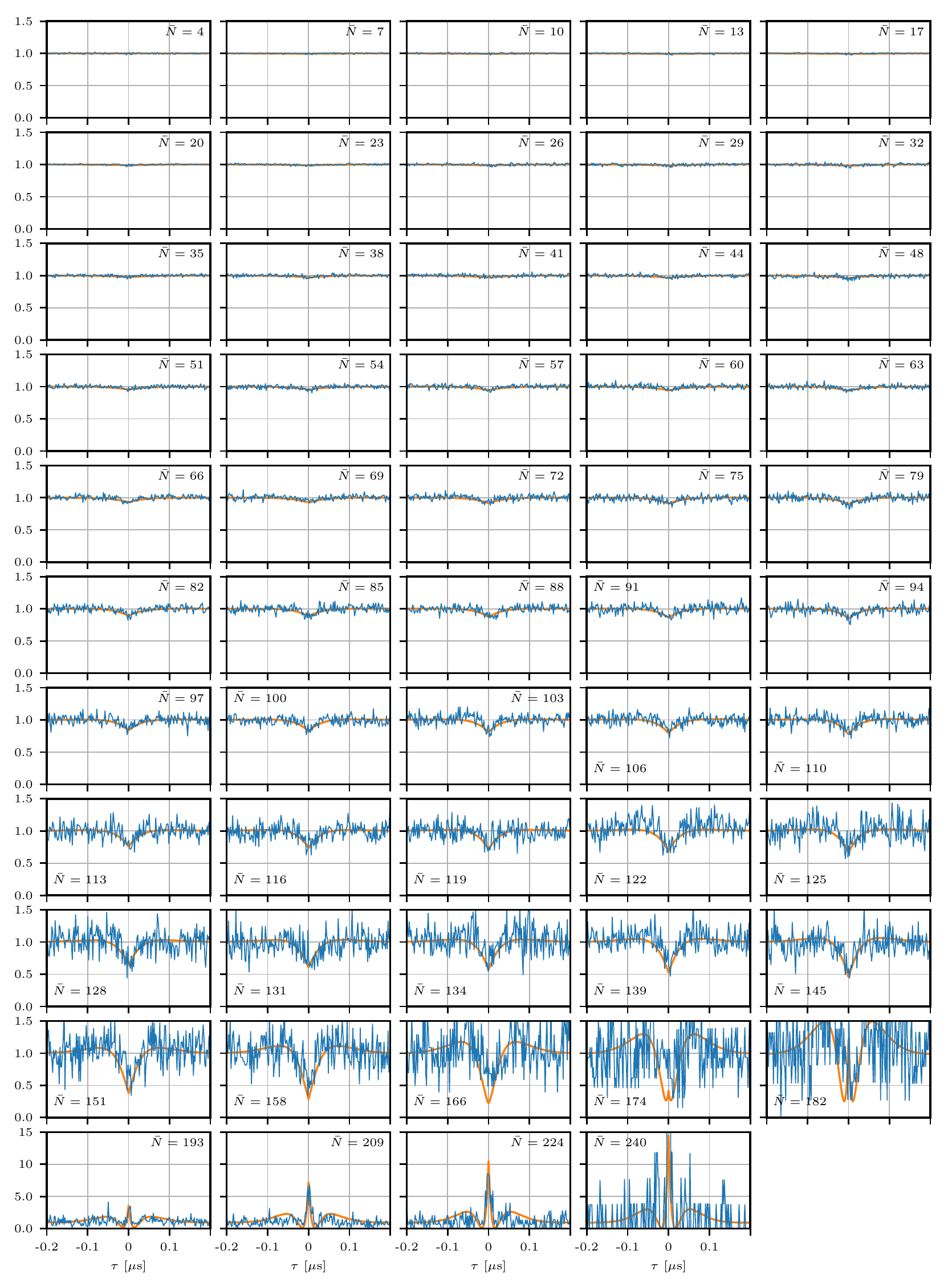}
	\caption{\textbf{Second order correlation function for different atom numbers} (blue) Measured correlation function for different atom numbers (see legend). (orange) Theoretically predicted correlation function, see main text.}
\end{figure*}

\end{document}